# Non- linearity Effect Analysis of Gausian Pulse Propagation In Optical Fiber

Yogesh Deo[1], Meenu Shrestha[1], Om Nath Acharya[1]
[1]Department of Electrical and Electronics
Engineering, Kathmandu University
(KU),Dhulikhel, Nepal,
Email:acharya.om@ku.edu.np

## Abstract

In this research, an extensive numerical analysis of nonlinear pulse propagation is conducted, focusing primarily on solving the nonlinear Schrödinger equation (NLSE) through the implementation of the split-step Fourier method. The study centers around the interaction between dispersive and nonlinear effects that occur concurrently within a nonlinear medium. These effects arise due to the medium's refractive index depending on the intensity of the propagating light, leading to the optical Kerr effect. This intensity-dependent refractive index contributes to a phenomenon known as self-phase modulation (SPM), which causes spectral broadening and temporal narrowing of optical pulses. At the same time, second-order group velocity dispersion (GVD) influences the pulse by causing it to broaden in time as different frequency components travel at slightly different velocities. While both of these effects are typically considered detrimental when acting alone. SPM leading to spectral distortion and GVD to pulse spreading. This project highlights how their interplay can result in a stabilizing effect on pulse propagation under specific conditions. A detailed exploration of GVD is followed by an in-depth discussion of SPM, analyzing how these effects influence the evolution of optical pulses. To illustrate these phenomena, Gaussian-shaped pulses are used as the initial input for numerical simulations carried out by solving the NLSE. The propagation dynamics of these pulses are observed and interpreted through the lens of the split-step Fourier method, a widely used numerical approach for solving the NLSE by separating the linear (dispersive) and nonlinear components over small spatial steps. Each stage of the algorithm is explained comprehensively, including the theoretical foundations and mathematical equations that define and guide the simulation process. This thorough approach allows for a clear understanding of how optical pulses behave in nonlinear dispersive media and underscores the significance of numerical methods in modeling complex physical systems.

# 1 Introduction

A fundamental goal of modeling fiber communications systems is to understand the physics of the system behavior and to develop computational tools to design systems and predict their performance. Transmission of data through a fiber-optic link unavoidably leads to bit errors due to various effects, the dominant of which are noise from optical amplifiers, fiber nonlinearity, polarization effects, and non-ideal transmitters and receivers.

Why is a careful analysis of nonlinear effects in optical fiber communications systems important? Nearly all modern systems operate in the linear propagation regime, in which the signal evolution is almost linear. However, there always exist small nonlinear interactions and small nonlinear signal distortions accumulated during transmission over hundreds and thousands of kilometers can lead to an increase in the error rate. Reducing the optical power decreases the importance of the nonlinear interactions, but it also decreases the signal-to-noise ratio. There exists an optimal power level at which the BER is minimal. Even if the power level is much lower than optimum, the accumulation of nonlinear distortions during transmission over hundreds or thousands of kilometers of fiber can introduce a significant system penalty [4].

If intensity of light goes high, refractive index of the core will be very much dependent on intensity, resulting in nonlinear effects in the material. These nonlinear effects are seen in the form of phase shifting of an optical pulse. Nonlinear dispersion produces a frequency chirp, is a signal whose frequency varies with time. All these nonlinear dispersions affect the transmission rate (bit rate) over fiber. Dispersion can be compensated using dispersion compensation techniques such as dispersion compensation fibers etc.

Researchers have found that Nonlinear Schrodinger Equation (NLSE) mapped to optical domain provides an adequate framework to describe optical signal propagation through nonlinear fiber for all the engineering applications found in practice. The most commonly used method in the fiber span simulation is split step method (SSM), which is based on the partition of the fiber spans into several spatial steps .

Project consists of fiber nonlinearities and execution of nonlinear Schrödinger equations using split step algorithm in MATLAB.

# 2 Literature Review

## 2.1 Optical fiber basics

An optical fiber is made up of a core, cladding and buffer coating where the core carries light pulses, the cladding reflects light pulses back into the core and the buffer coating is used to prevent core and cladding from being damaged. Optical fiber is advantageous over the metal wires as it can transmit the data over longer distance with less loss of data sent, higher bandwidth (data rates of 10-40 Gbps) and high resistance to electromagnetic noise. The basic structure of an optical fiber is shown below.[1]

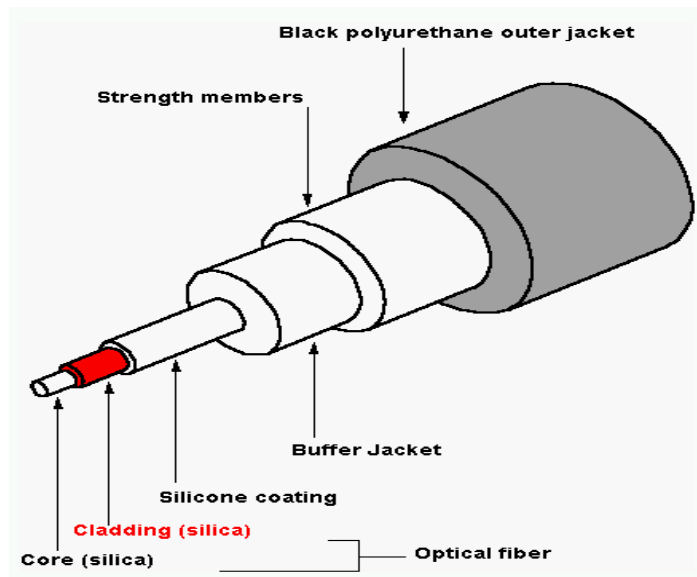

Fig 2.1 Optical fiber

The diameter of the core is smaller than cladding and buffer coating as shown in fig 2.1. The basic parameter of optical fiber is "refractive index". Refractive index, defined as ratio of speed of light wave and phase velocity of the wave in the medium. It varies with respect to different mediums (air, water etc).

## 2.2  Optical fiber transmission link

The key elements of optical transmission link are transmitter block, optical channel and receiver block. Below is shown block diagram. [1] Transmitter consists of optical source, modulation and drive circuit.

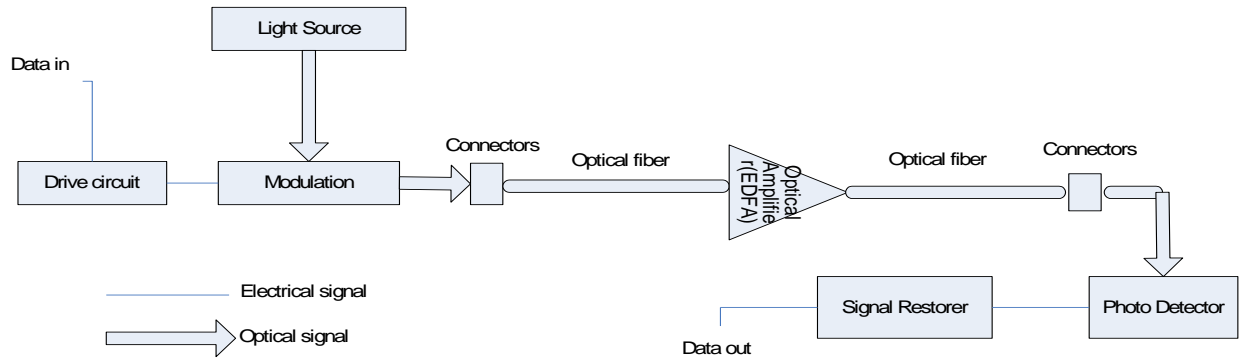

Fig 2.8 Optical fiber transmission link

In the transmitter block, data in is generated by random generator and sent to drive circuit (electrical generator) to represent the signal in Return to Zero (RZ) or Non Return to Zero (NRZ) format. The output of the drive circuit and output of the light source is send to modulator block to modulate the optical signal (from light source). Connectors are used to provide the interface between transmitter block to transmission medium (optical channel) and transmission medium to receiver block. Optical channel comprises of optical fiber as transmission medium and optical amplifier. Optical amplifier is used to boost the optical signal without converting to electrical signal. These amplifiers are doped to ensure reliable data with the increase of transmission distance.  Photo detectors and signal restorer are in the receiver block. Photo detectors (mostly photodiodes) are used to sense the light signals and convert them to electrical signal (current/voltage). Signal restorer is used to extract the signal from noise-induced signal.

## 2.3  Dispersion

If a light signal is transmitted over a long haul optical fiber, its power is dispersed with respect to time which widens shape of the pulses in the signal with time. This is called as "Dispersion" (pulse broadening) of the signal. Below is the visual representation of widening shape of the pulse when transmitted through fiber.

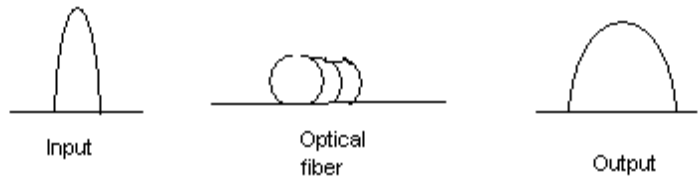

Fig 2.7 Dispersion in fiber

Signal dispersion is seen due to multiple modes in the fiber, fiber material, waveguide of fiber and nonlinearities in fiber.

### 2.3.1 Modal Dispersion

Modal dispersion can be seen in multimode fibers as it's a result of differences in group velocities of different modes in the multimode fiber. The group velocity is defined as velocity at which data is transmitted along the wave and is given by expression

$$V_g = \frac{\partial \omega}{\partial \beta}$$

$$\omega = \frac{2\pi c}{\lambda}$$

Where $V_g$ = group velocity of a wave,

$w$ = angular frequency,

$\beta$ = propagation constant,

c = velocity of light (m),

$\lambda$ = wavelength of the fiber (nm)

Modal dispersion also referred as "Intermodal dispersion" as it is the distortion caused by differences in delay times of the all modes. Time delay of a mode is defined as ratio of length of fiber and group velocity of the same mode and given by expression [2]

$$\tau_q = \frac{L}{V_g}$$

Where $\tau_q$ = time delay of mode "q", $V_g$ = group velocity of mode "q", L= length of fiber

Dispersed RMS pulse width can be calculated with an estimated expression

$$\sigma_\tau \approx \frac{1}{2}\left[\frac{L}{V_{min}} - \frac{L}{V_{max}}\right]$$

$V_{min}$ and $V_{max}$ are smallest and largest group velocities and L is the length of the fiber.

Polarization mode dispersion is a form of modal dispersion, in which different polarizations of light in waveguide travel at various speeds because of the imperfections and irregularities causing pulse spreading of the optical pulse.

### 2.3.2 Material and Waveguide dispersion

Another kind of dispersion is due to the material of the fiber is known as "Material dispersion". Optical pulse is combination of spectral components, which operates at their respected wavelengths for a given mode. These spectral components propagate with different speeds depending on their wavelengths. As refractive index varies with the optical wavelengths, material dispersion occurs. RMS value of width of pulse spread by material dispersion is given by

$$\sigma_\tau = |D_\lambda|\sigma_\lambda L$$

$$D_\lambda = -\frac{\lambda_0}{c_0}\frac{d^2 n}{d\lambda_0^2}$$

Where $\sigma_\tau$ = width of pulse spread, L= length of the fiber, n= Refractive index of medium, $\sigma_\lambda$ = width of the source pulse, $D_\lambda$ = Material dispersion (ps/km-nm)

Waveguide dispersion is another type of dispersion seen in the fiber. It occurs when a signal travels in a waveguide, which has irregular geometry and structure and refractive index is independent of wavelength [4]. Expression for width of pulse spread and waveguide dispersion is

$$\sigma_\tau = |D_w|\sigma_\lambda L$$

$$D_w = \frac{d}{d\lambda_0}\left(\frac{1}{\upsilon}\right) = -\frac{\omega}{\lambda_0}\frac{d}{d\omega}\left(\frac{1}{\upsilon}\right)$$

$D_w$ = Waveguide dispersion, $\lambda_0$ = Wavelength of the fiber, $\upsilon$ = frequency of signal

Material and waveguide dispersion together referred as "Chromatic dispersion". Chromatic dispersion is an intramodal dispersion because pulse broadening is related with only single mode of the fiber.

### 2.3.3 Nonlinear dispersion

If intensity of light goes high, refractive index of the core will be very much dependent on intensity, resulting in nonlinear effects in the material. These nonlinear effects are seen in the form of phase shifting of an optical pulse. Nonlinear dispersion produces a frequency chirp, is a signal whose frequency varies with time. [2]

All these dispersions affect the transmission rate (bit rate) over fiber. Dispersion can be compensated using dispersion compensation techniques such as dispersion compensation fibers etc. [2]

## 2.4  Nonlinear effects

In an optical system, obtaining the maximum transmission rate is possible by merging TDM and WDM concepts as it depends on temporal and spectral characteristics of light. However, TDM has limiting factors in terms of chromatic and polarization mode dispersion whereas WDM has limiting factors in terms of the non-linear effects for the transmission performances in the fiber.

Optimized maximum transmission capacity depends on a few factors such as dispersion, signal power and fiber length. [7]

Nonlinearity in optical fiber is caused due to high intensity of light in the core. There are two possibilities for the occurrence of nonlinear effects in optical fiber. They are due to inelastic-scattering phenomenon or due to change in the refractive index of the medium related with intensity of light.[5]

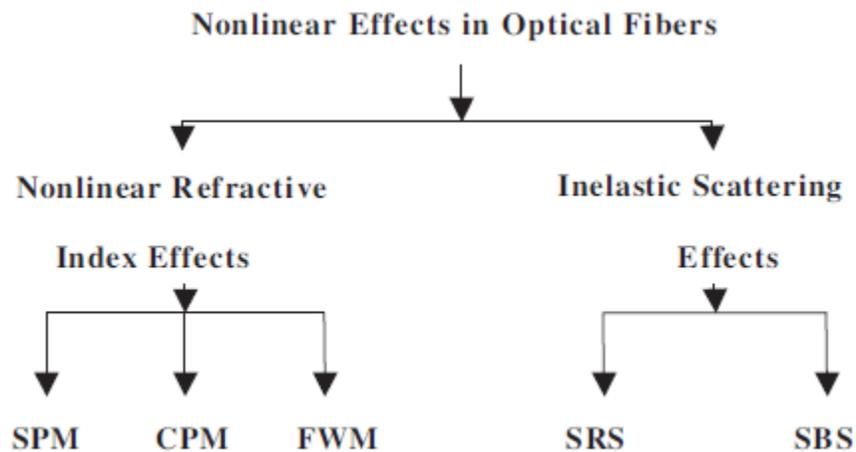

Fig 3.1 Types of nonlinear effects

Non-linear effects have their own advantages and disadvantages. These effects limit transmission capacity and can be overcome by using large core fibers, reverse dispersion fibers etc.[7] On the other side, these effects can be used to improve the performance of the system by wavelength conversion.

### 2.4.1 Non Linear Refractive Index

Nonlinear effects in fiber optic are mostly generated due to nonlinear refraction. Refractive index generally defines the density of the propagating medium thereby describing how light would propagate and at what speed. When light of very high intensity is launched into the fiber, an additional variable contributes to the total refractive index. The total refractive index is defined as follows:

$$n = n_0 + n_2|I|^2$$

Here, $n_0$ is the material refractive index and it is assumed constant for a certain frequency. The second term $n_2|I|^2$ is defined as the nonlinear refractive index where $n_2$ is nonlinear index coefficient.

When the light pulses are having short wavelengths and very high intensities (such as output of a laser) may vary a refractive index of a medium which as a result may give rise to nonlinear optics. If the refractive index of the medium varies nonlinearly with the field (linearly with the intensity), then it is known as the optical Kerr effect [3] and results in phenomena such as self-focusing and self-phase modulation. If the index varies linearly with the field then it is known as the Pockels effect.

The value of $n_2$ is affected by the experimental technique used to measure it. The reason is that two other mechanisms which is related to molecular motion (the Raman effect) and excitation of acoustic waves through electrostriction (Brillouin scattering), also contribute to $n_2$. However, their relative contributions depend on whether the pulse width is longer or shorter than the response time associated with the corresponding process.

### **Kerr effect**

Nonlinearity in the refractive index is known as Kerr nonlinearity, producing a carrier-induced phase modulation of the propagating signal, which is called the Kerr effect. Refractive index is a function of electric field $E$ denoted by n(E). The function of refractive index is evaluated by Taylor's series given by

$$n(E) = n + \frac{dn}{dE} + \frac{1}{2}\frac{d^2n}{dE^2}E^2 + ........$$

For symmetry material, first order term of E is zero as refractive index is an even function. Second order is consider to study the Kerr effect and given by an expression

$$n(E) = n - \frac{n^3\xi}{2}E^2$$

Where Kerr coefficient is given by

$$\xi = \frac{1}{n^3} \frac{d^2 n}{dE^2}$$

n= refractive index of the medium

Typical value of Kerr coefficient is $10^{-18}$ to $10^{-14}$ $\frac{m^2}{V^2}$

Kerr nonlinearity further gives rise to three different effects based on input signal power such as self-phase modulation, cross phase modulation and four wave mixing.

### 2.4.1.1 Self-Phase Modulation (SPM)

Nonlinear effect depends on intensity of light and refractive index. Input pulse travels through fiber, which has high intensity of light inside core resulting in higher refractive index. Signal intensity changes with respect to time leads to variations in refractive index with time, which is similar to intensity dependent refractive index.These variations in refractive index resulting in time dependent phase changes. These phase changes are the same as optical signal change with time, so the name Self Phase Modulation (SPM). Different parts of the input pulse change the phase of the signal randomly, resulting a frequency chirp, which is defined as a signal whose frequency varies with time (increases-up chirp or decreases-down chirp). These variations in frequencies cause pulse broadening, which can be seen significantly high in the systems with high transmission power because the transmission power is directly dependent on frequency chirp. [5]

Phase shift by field over fiber length is given by

$$\varphi = \frac{2\pi n L}{\lambda}$$

Where n= refractive index of the medium, L= length of the fiber, $\lambda$ = Wavelength of the optical pulse, nL= Optical path length

In SPM, pulse broadening is seen in the time domain and spectral characteristics are unaltered. Chirp produced by SPM is used to reduce the effects of dispersion caused by pulse broadening.

These effects depend mostly on Input power of the signal transmitted, which can be used as a threshold condition for the frequency chirp to occur. SPM is major limitations in single channel systems. [5]

### 2.4.1.2 Cross Phase Modulation (CPM)

Cross phase modulation is similar nonlinear effect to self phase modulation except it occurs when there are two or more optical signals propagate through fiber. The distortion of signal and pulse broadening will be asymmetric because of more than one signal is propagating and depends on intensity of the propagating pulse and co-propagating pulses. "CPM converts power fluctuations in a particular wavelength channel to phase fluctuations in other co-propagating channels". [5]

Expression for phase shift caused by nonlinear effect is given by

$$\phi_{nl}^{i} = k_{nl} L_{eff} \left( P_i + 2 \sum_{n \neq i}^{N} P_n \right) \quad [5]$$

N= N-channel transmission system, n= 1, 2, 3…...N, $L_{eff}$ = Effective length of link $k_{nl}$ = Propagation constant

CPM is better over SPM for propagating transmission capacity twice but it is advantageous only when all the propagating signal are superimposed with each other for every time slice. If the signals are not superimposed, it can lead to severe damage to the system performance when compared to SPM. [5]

### 2.4.1.3 Four Wave Mixing (FWM)

In this nonlinear effect, three optical fields when propagated through fiber will give rise to a new optical field, which depends on the three optical fields, which it was originated from. The frequency of the new optical field is given by [5]

$$\omega_4 = \omega_1 \pm \omega_2 \pm \omega_3$$

ω1, ω2, ω3 are the frequencies of three original optical fields

FWM is not dependent on bit rate as the other two nonlinear effects; instead they depend on channel spacing and dispersion of fiber. Dispersion depends on wavelength, so newly generated optical wave and reference signal wave have different group velocities. With different group velocities, phase matching is not possible, which decrease power transfer to new optical wave. At the same time, if the newly generated wave and original wave has same wavelength, results in interference. The interference of the signals decreases signal to noise ratio. Whenever we observe different group velocities then FWM effect decreases, channel spacing increases and so does dispersion. [5]

In conclusion, nonlinear effects degrade the performance of fiber optic systems. They provide gain to some channels by increasing transmission rate but at an expense of depleting power. SPM and CPM affect only phase of the signals and can cause spectral broadening, which leads to increased dispersion. The nonlinear effects depend on transmission length, intensity. [5]

## 2.5  Nonlinear Schrodinger Equation

Nonlinear Schrodinger equations is used to analyze the nonlinear behavior of signals propagating in an optical fiber as it includes the physical effects, dispersion and nonlinearity, of the propagating signal. [6]

The basic form of NLSE is

$$\frac{\partial A}{\partial z} + \frac{i}{2}\beta_2 \frac{\partial^2 A}{\partial t^2} = i\gamma |A|^2 A - \frac{\alpha}{2} A$$

where $A = A(z,t)$ is a amplitude of a Gaussian input pulse and is associated with the electric field E of an optical signal in a fiber, $\beta_2$ = Group velocity dispersion coefficient, $\alpha$ = fiber losses db/km, $\gamma$ = Nonlinear coefficient.

Propagation of Gaussian pulse(with chirp) in fiber is consider by taking initial amplitude of that pulse, which is given by

$$A(0,t) = A_0 \exp\left[-\frac{1+iC}{2}\left(\frac{t}{T_0}\right)^2\right]$$

Where $A_0$ = peak amplitude, C= Chirp factor, $T_0$ and t are the input pulse width and time period of the wave travelled, Nonlinear coefficient is calculated using an expression

$$\gamma = \frac{n_2 \omega}{c A_{eff}}$$

Where $n_2$ = refractive index of cladding and depends on refractive index of core, $\omega$ = angular frequency, $c$ = speed of light, $A_{eff}$ = core area of fiber.

# 3   Methodology

## 3.1   Description of method

The numerical analysis of these nonlinear effects is done by Nonlinear Schrödinger equation. The equation is solved using an algorithm called "Split-step algorithm".

### 3.1.1   Split Step Algorithm

The split step algorithm is a pseudo-spectral numerical method for solving partial differential equations such as the nonlinear Schrödinger equation. Split step algorithm separates linear and nonlinear parts of the equation as shown below and solves it separately. It is applied because of greater computation speed and increased accuracy compared to other numerical techniques.

Dispersion and nonlinear effects act simultaneously on propagating pulses during nonlinear pulse propagation in optical fibers. However, analytic solution cannot be employed to solve the NLSE with both dispersive and nonlinear terms present. Hence the numerical split step fourier method is utilized, which breaks the entire length of the fiber into small step sizes of length 'h' and then solves the nonlinear Schrödinger equation by splitting it into two halves , the linear part (dispersive part) and the nonlinear part over z to z + h [7].

Each part is solved individually and then combined together afterwards to obtain the aggregate output of the traversed pulse. It solves the linear dispersive part first, in the fourier domain using the fast fourier transforms and then inverse fourier transforms to the time domain where it solves the equation for the nonlinear term before combining them. The process is repeated over the entire span of the fiber to approximate nonlinear pulse propagation. The equations describing them are offered below [7].

The nonlinear simplified Schrödinger equation is given by

$$\frac{\partial A}{\partial Z} + \frac{i\beta_2}{2}\frac{\partial^2 A}{\partial t^2} = i\gamma \left|A^2\right| A - \frac{\alpha}{2}A$$

We can rewrite the above equation as

$$\frac{\partial A}{\partial Z} = -\frac{i\beta_2}{2}\frac{\partial^2 A}{\partial t^2} - \frac{\alpha}{2}A + i\gamma|A^2|A$$

Representation of above equation after dividing into linear and nonlinear parts is

$$\frac{\partial A}{\partial Z} = -\frac{i\beta_2}{2}\frac{\partial^2 A}{\partial t^2} - \frac{\alpha}{2}A + i\gamma|A^2|A = [D+N]A$$

When γ=0, results in linear part of nonlinear Schrödinger equation

$$\frac{\partial A_D}{\partial Z} = -\frac{i\beta_2}{2}\frac{\partial^2 A}{\partial t^2} - \frac{\alpha}{2}A = DA$$

Consider α=0, $\beta_2$=0, results in nonlinear part of nonlinear Schrödinger equation

$$\frac{\partial A_N}{\partial Z} = i\gamma|A^2|A = NA$$

Added a small step "h" simulation parameter is added to separate the linear and nonlinear terms of the equation with minimal error. If we solve nonlinear part of equation in time domain will result as [10]

$$A_N(t, Z+h) = \exp\left[i\gamma|A|^2 h\right]A(t,Z)$$

In the same way, solving the equation of linear part gives us

$$A(\omega, Z+h) = \exp\left[\frac{i\beta_2}{2}(\omega)^2 h - \frac{\alpha}{2}h\right]A_D(\omega, Z)$$

The linear function is solved in Matlab and then took inverse Fourier transform of the linear function and multiplied with the nonlinear function as both are in time domain. A Fourier transform is taken to generate the spectrum for combined equation and inverse Fourier transform ensures the propagation of signal.

# 4 Simulation

## 4.1 Simulation parameters

**1$^{st}$ case**: For the ideal case, the parameters are

$P_i$ (input power)= 3.98mw

Fiber losses in db/km= 0

Time period of input pulse= 200ps

area of effective core = 67.56

Chirp factor =0

Wavelength= 1550nm

dispersion coefficient= -10 ps/nm/km and

length of the fiber = 100km

Time period of upper period= - 4096ps

Time period of lower period=4095ps

Time interval= 1ps

**2$^{nd}$ case**: To show pulse evolution and pulse broadening ratio with different chirps, fiber non-linearity and dispersion coefficient.

Pi=0.00064mw or (0.001, 0.005, 0.0001)

gamma= 0.003 or (0.003, 0.006, 0.009, 0.011 and 0.001 /W/m.)

Group velocity dispersion coefficient, $\beta_2$ = = - 20*10^(-27) or (- 20*10^(-27) , -40*10^(-27), -60*10^(-27) and -100*10^(-27) )

wavelength =1550nm

Chirp factor = 0 and -2

time period of input pulse is 125ps

length of the fiber = 15km and other parameters remain the same.

## 4.2 Simulation Graphs and Analysis

**For the 1st case ( Ideal case):**

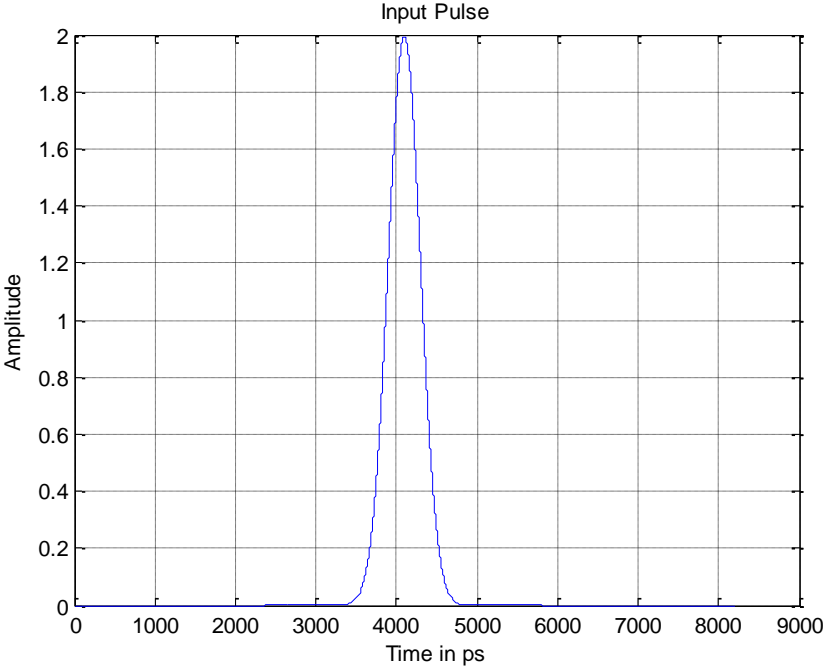

Figure 4-1 Input pulse ( input power= 3.98mW)

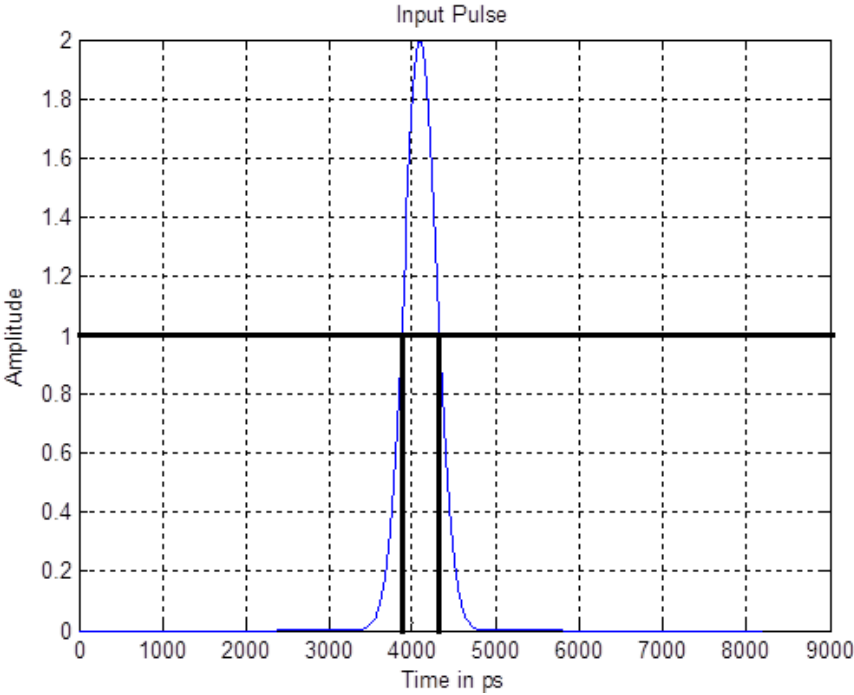

Figure 4-2 FWHM points on input pulse

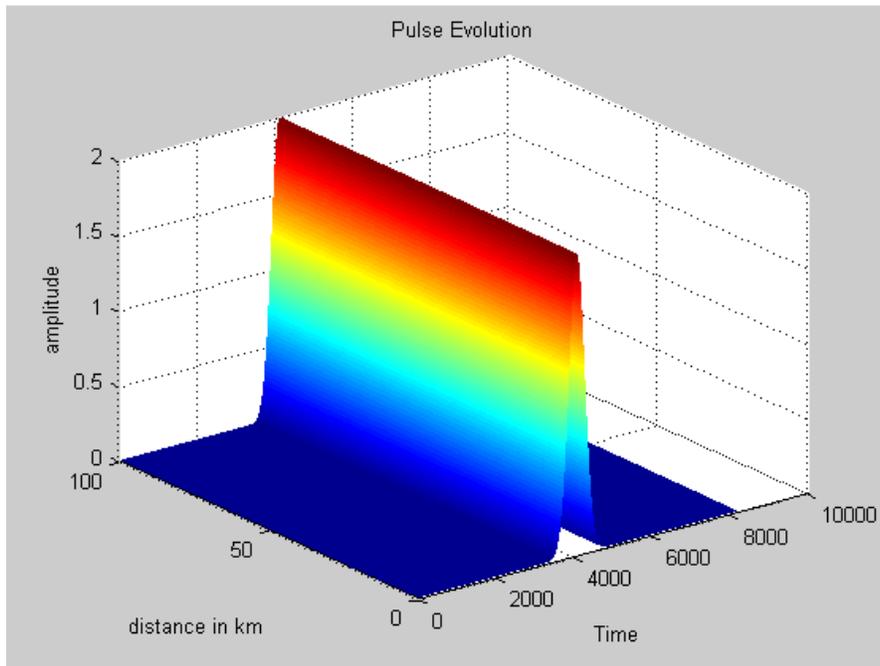

Figure 4-3 Output spectrum of input pulse (ideal)

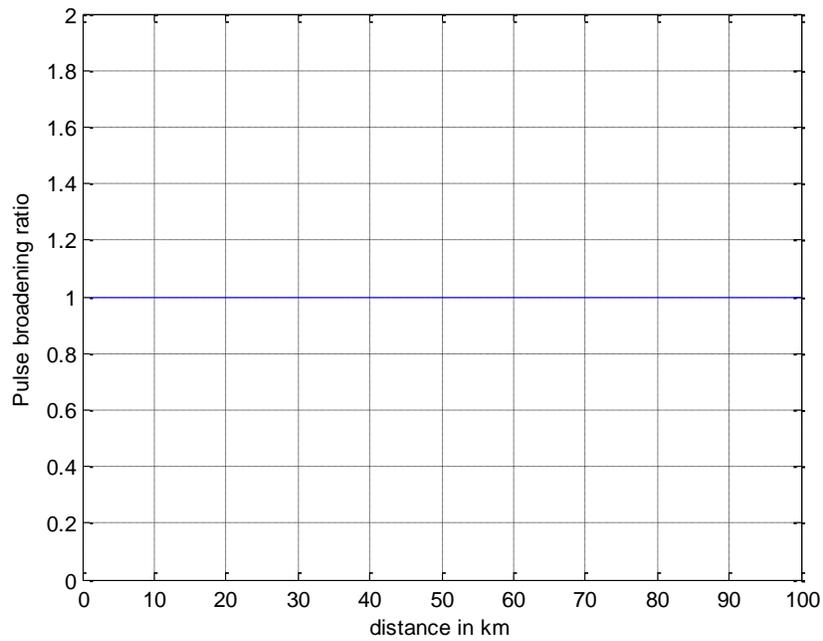

Figure 4-4 Pulse broadening plot

## **Analysis**

Fig 4.1 indicates input pulse, Fig 4.2 shows us FWHM (Full Width at Half Maximum, defined as width of frequency range where at least signal power is half the maximum) points on input pulse.

FWHM points are observed at half of the power, if amplitude is calculated with power of signal, if amplitude is calculated with voltage, then FWHM points are observed at 0.707*Voltage .

Pulse broadening ratio is defined as ratio of output FWHM to input FWHM. With the help of FWHM's generated from the code, pulse broadening ratio is plotted. The pulse broadening ratio plot shown in Fig 4.3 explains how the input pulse is broadening with respect to distance travelled.

Fig 4.4 shows us the spectral output pulse waveform (pulse broadening is zero for ideal case). Ideally, there is no pulse broadening, when input pulse transmitted through zero dispersion and zero chirp factor in the fiber calculated by numerical analysis but when an input pulse is sent through the fiber, simulation results show dispersion. From Fig 4.4, when there is no pulse broadening the received signal will be replicate of input signal considering zero losses.

**For the 2$^{nd}$ case : 1$^{st}$ part ( with chirp=0)**

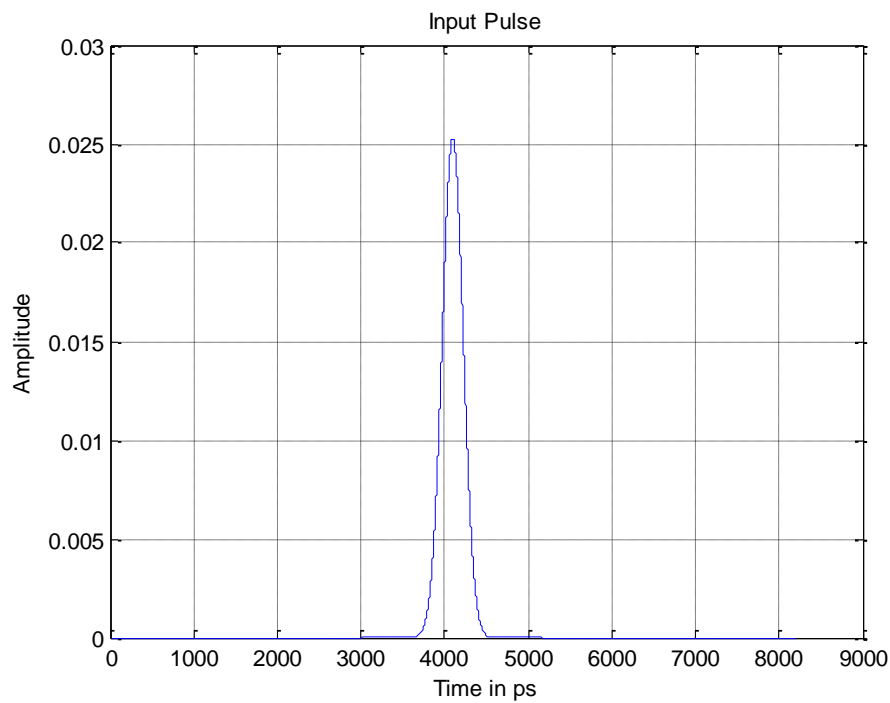

Figure 4-5 Input pulse ( input power= 0.0064mW)

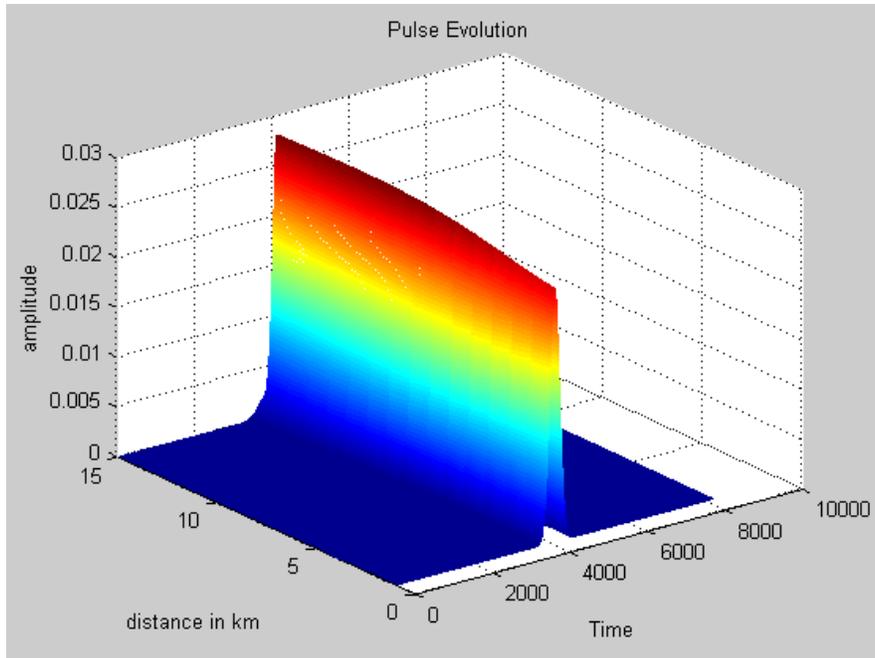

**Figure 4-6 Output spectrum(mesh) of input pulse (with chirp=0)**

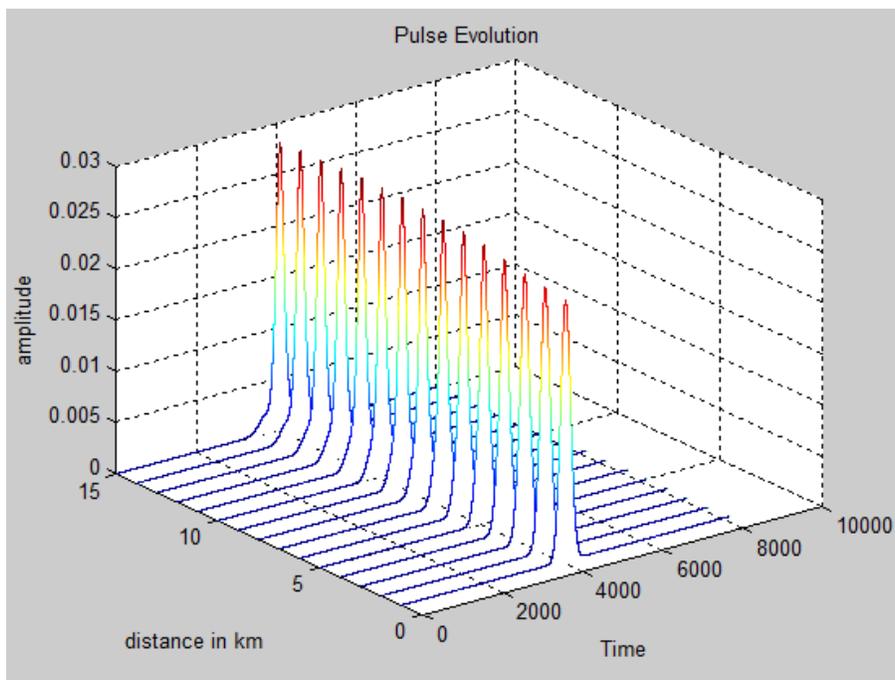

**Figure 4-7 Output spectrum( waterfall) of input pulse (with chirp=0)**

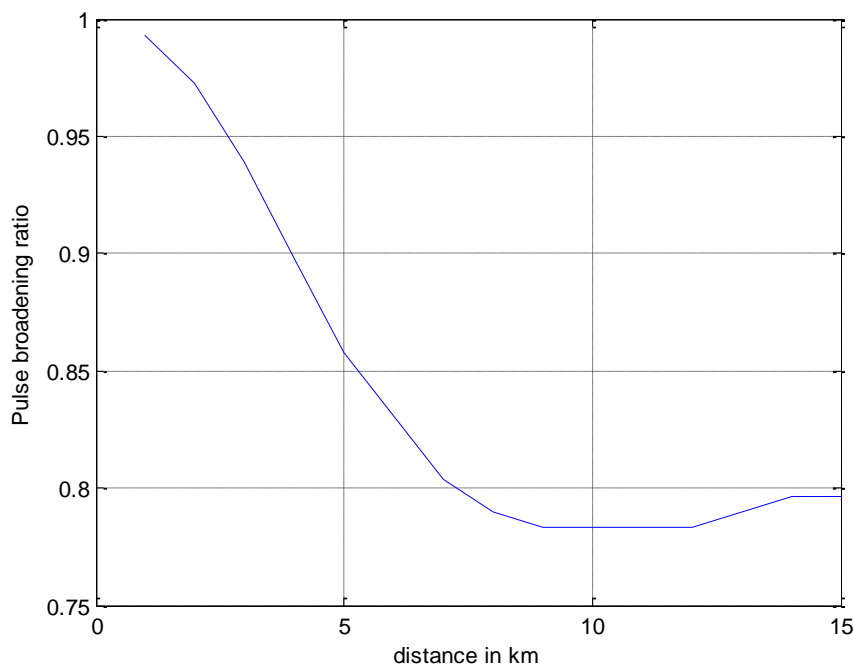

**Figure 4-8 Pulse broadening plot (with chirp =0)**

**Analysis:**

Here, both GVD and SPM act simultaneously on the propagating Gaussian pulse with no initial chirp. As can be seen from the evolution graph in Figure 4-7, the pulse shrinks initially for a very small period of propagating length. After that the broadening ratio reaches a constant value and a stable pulse is seemed to propagate as can be seen from pulse broadening ratio curve in Figure 4-8. GVD acting individually results in the pulse to spread gradually before it loses shape. SPM acting individually results in the narrowing of pulses and losing its intended shape.

The combined effect of GVD and SPM leads to the eventual generation of constant pulse propagation. Initially, the positive chirp induced by SPM seems to dominate the negative chirp caused by the GVD which accounts for the narrowing of the propagating pulse in the early steps. But after a certain period, GVD effect becomes more prominent and it catches up to the SPM effect. At one point, GVD induced negative chirp balances the SPM induced positive chirp, and they cancel each other out and propagates at a narrower pulse width emulating hyperbolic secant pulse propagation.

# For the 2nd case : 2nd part ( with chirp= -2)

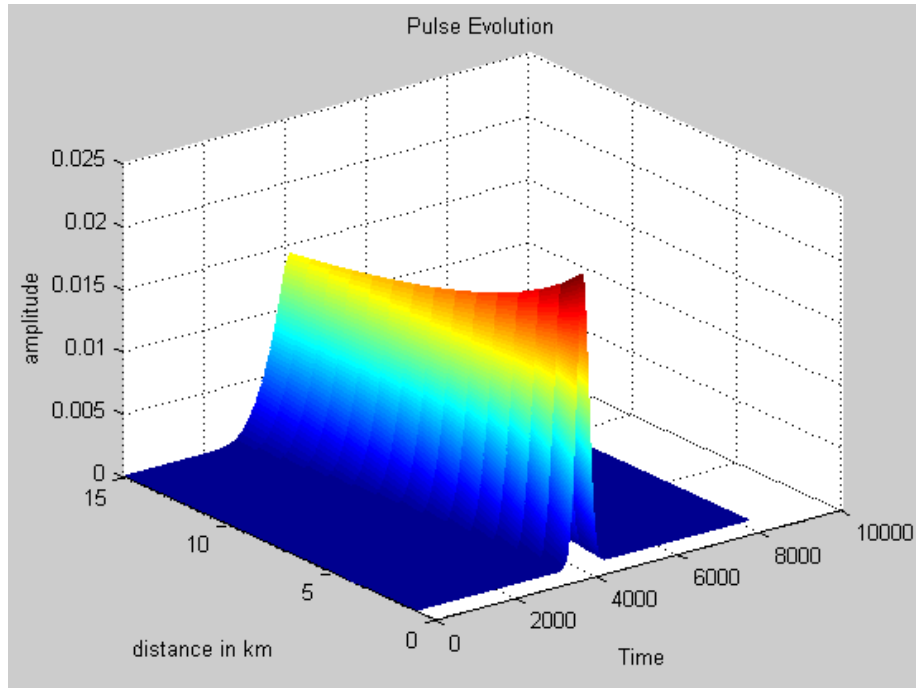

Figure 4-9 Output spectrum ( mesh) of input pulse (with dispersion),chirp=-2

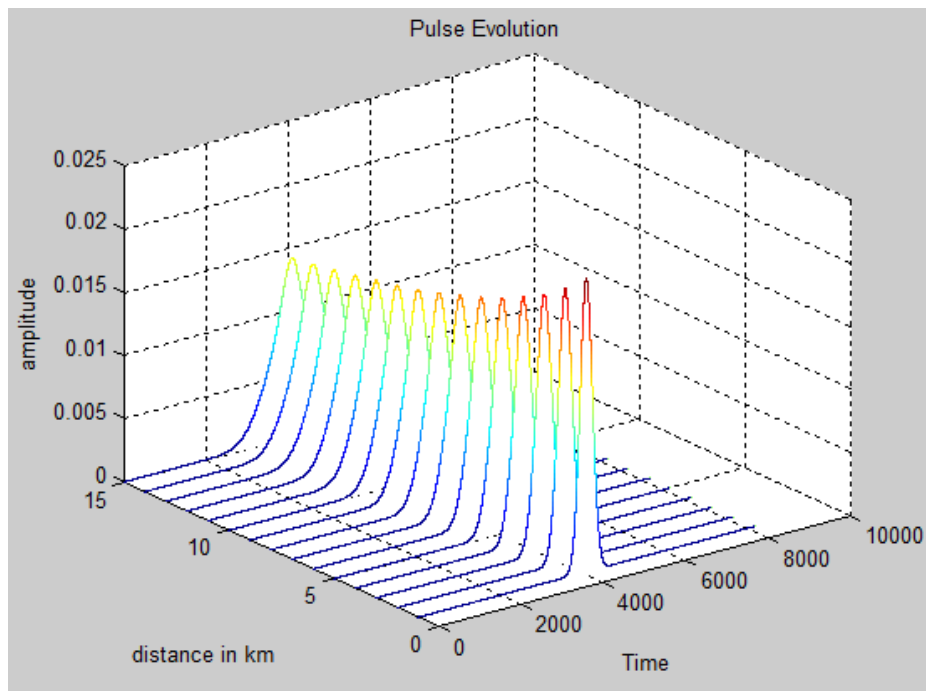

Figure 4-10 Output spectrum ( waterfall ) of input pulse (with dispersion),chirp=-2

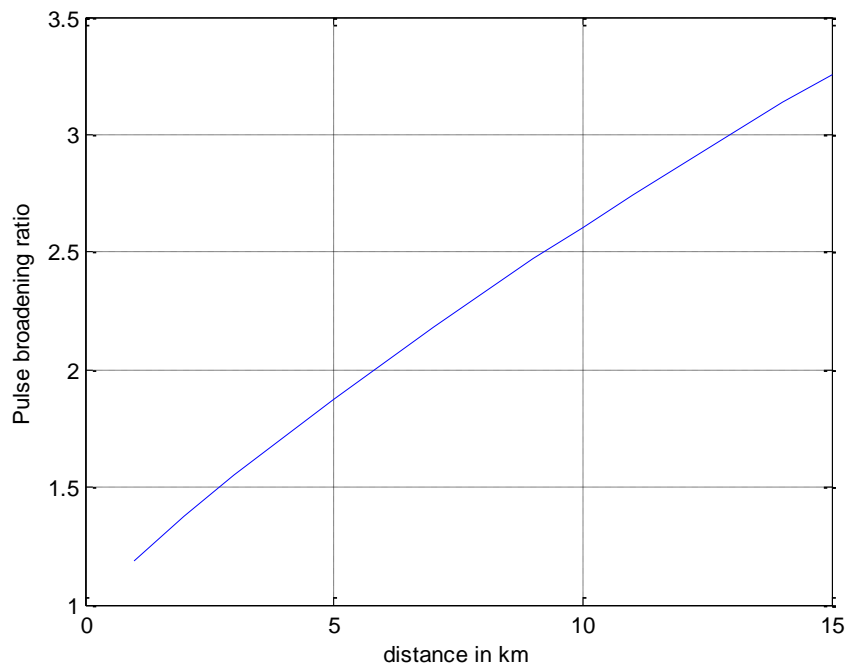

**Figure 4-11 Pulse broadening plot (with dispersion), chirp=-2**

**Analysis:**

Here, both GVD and SPM act simultaneously on the gaussian pulse with initial negative chirp. From Figure 4.10 we observe that the pulse width broadens as it propagates. From the pulse broadening ratio curve, we see that the pulse broadens from the very start and continues to broaden. It is also obvious that the rate at which the pulse broadens is higher than in previous cases.

As before, we concern ourselves with anomalous dispersion region where $\beta_2 < 0$. Therefore for C= -2 we have $\beta_2 C > 0$ which as explained previously means that the direction of dispersion induced chirp is in the same direction to that of the initial chirp value. The negative input chirp C adds to the negative induced chirp of the GVD. At the same time the positive chirp caused by SPM is reduced due to deduction from the negative input chirp. Hence, as the increasing GVD effect causes dispersion of the pulse at an enlarged rate, the curtailing of the SPM effect results in the reduction of the rate at which the pulse narrows. The net effect is that the rate of dispersion is much higher than the rate of pulse narrowing which explains why the pulse broadening ratio increases at a much higher rate than previous cases.

Also, unlike in the case of chirp 0, the impact of the initial chirp parameter is greater if the magnitude of the chirp is more than a certain value. In such a case, the GVD and SPM do not cancel each other further down the length of the propagating medium; instead the pulse continues to broaden indefinitely. This value of the initial chirp where GVD effect grows and eventually dominates pulse propagation is the critical chirp. Beyond such a value, the transformation of the Gaussian pulse into a constant width pulse is impossible.

# 5   Conclusion

In this project numerical analysis of nonlinear pulse propagation is carried out mainly by solving the nonlinear Schrodinger equation using the split step algorithm. In a nonlinear media, dispersive effects exist simultaneously with nonlinear effects. Refractive index dependence on intensity results in optical Kerr effect which causes narrowing of transmitted pulses by inducing self-phase modulation while second order group velocity dispersion causes the pulses to spread. These individually detrimental effects are shown to combine beneficially for propagation of pulses here. Gaussian pulse is studied and propagated by using them as input in to the nonlinear Schrodinger equation. The split step algorithm is described in depth. Explanation of each step is included along with the relevant equations defining these steps. Initially, we discussed the theoretical analysis and comparison of nonlinear effects then practically proved the theoretical analysis of nonlinear effects by simulating nonlinear Schrödinger equation using Split step algorithm to analyze the effects of nonlinearity in fiber, which is coded and optimized in Matlab.

# 6 Appendix

**Matlab code**

The Matlab code was developed to observe the numerical analysis through simulations. Reference code to develop this code is enlisted in the reference section [7]. For the value of the parameters please refer the Simulation parameter subsection of chapter 4.

**For Case1( ideal case)**

```matlab
clc;
close all;
clear all;
q=1;

%% User data
Pi=input('Enter the value of input power in mW   ')
alpha=input('Enter the value of fiber loss in db/km   ')
t=input('Enter the value of input pulse width in seconds   ')
tau=input('Enter time period with upper(U), lower(L) and interval between upper and  lower interval(I) in this format L:I:U')
dt=input('enter intervel of period')
Area=input('Enter the area of effective core in m^2   ')
C=input('Enter the value of chirp factor    ')
Lamda=input('Enter the wavelength in meters    ')
D=input(' Enter the value of dispersion coefficient ps/nm/km    ')
s=input('Enter the length of fiber in Km    ')
disp('Fiber losses in /km    ')

%%
A=alpha/(4.343)
c=3*1e8;
n1=1.48;
pi=3.1415926535;
delta=(n1*2*pi*s)/Lamda
n2=7e-10
n2=n1*(1-delta)
f=c/Lamda;
omega=2*pi*f;
a=n2*omega;
b=c*Area;
gamma=a/b %calculation of nonlinear coefficient
B2=-(Lamda^2*D)/(2*pi*c) % calculation of group velocity dispersion coefficient
L=(t^2)/(abs(B2)) %dispersion length

Ao=sqrt(Pi) % amplitude of an input gaussian pulse

 h=1000   % small step as defined by  split step algorithm

for ii=0.1:0.1:(s/10) %different fiber lengths
```

```matlab
X=ii*L;
    At=Ao*exp(-((1+i*(C))/2)*(tau/t).^2); % generation of an gaussian input pulse
    figure(5)
   plot(abs(At)); % graph of input pulse
   title('Input Pulse'); xlabel('Time in ps'); ylabel('Amplitude');
grid on;
hold off;
l=max(size(At));

fwhmi=find(abs(At)>abs(max(At)/2));
fwhmi=length(fwhmi);
dw=1/l/dt*2*pi;
w=(-1*l/2:1:l/2-1)*dw;
At=fftshift(At);
 w=fftshift(w);
spec=fft(fftshift(At)); %generating a pulse spectrum
for j=h:h:X
spec=spec.*exp(-A*(h)+i*B2/2*w.^2*(h)) ; % calculation of linear part of NLSE
f=ifft(spec);
f=f.*exp(i*gamma*((abs(f)).^2)*(h)); %calculating nonlinear part of NLSE
spec=fft(f);
spec=spec.*exp(-A*(h/2)+i*B2/2*w.^2*(h/2)) ; % NLSE calculation
end
f=ifft(spec); % pulse propagation
output_pulse(q,:)=abs(f); %preserving output pulse
fwhmo=find(abs(f)>abs(max(f)/2));
fwhmo=length(fwhmo);% finding the Full width half maximumof output signal
pbratio(q)=fwhmo/fwhmi; %finding pulse brodening ratio
q=q+1;
end
figure(6);
mesh(output_pulse(1:1:q-1,:)); %output spectrum
title('Pulse Evolution');
xlabel('Time'); ylabel('distance in km'); zlabel('amplitude');
figure(7)
plot(pbratio(1:1:q-1),'b'); %pulse broadening plot
xlabel('distance in km');
ylabel('Pulse broadening ratio');
grid on;
hold on;
```

**For Case 2:**

```matlab
clc;
close all;
clear all;
q=1;

%% User data
Pi=input('Enter the value of input power in mW   ')
alpha=input('Enter the value of fiber loss in db/km   ')
t=input('Enter the value of input pulse width in seconds   ')
```

```matlab
tau=input('Enter time period with upper(U), lower(L) and interval between upper and  lower interval(I) in this format L:I:U')
dt=input('enter interval of period')
C=input('Enter the value of chirp factor   ')
Lamda=input('Enter the wavelength in meters   ')
s=input('Enter the length of fiber in Km    ')
disp('Fiber losses in /km    ')

%%
A=alpha/(4.343)
c=3*1e8;
n1=1.48;
pi=3.1415926535;
delta=(n1*2*pi*s)/Lamda
n2=7e-10
n2=n1*(1-delta)
f=c/Lamda;
gamma=0.003; nonlinear coefficient
B2= -20*10^-27; %group velocity dispersion
L=(t^2)/(abs(B2)) %dispersion length

Ao=sqrt(Pi) % amplitude of an input gaussian pulse

 h=1000  % small step as defined by  split step algorithm

for ii=0.1:0.1:(s/10) %different fiber lengths
X=ii*L;
    At=Ao*exp(-((1+i*(C))/2)*(tau/t).^2); % generation of an gaussian input pulse
    figure(1)
   plot(abs(At)); % graph of input pulse
   title('Input Pulse'); xlabel('Time in ps'); ylabel('Amplitude');
grid on;
hold off;
l=max(size(At));

fwhmi=find(abs(At)>abs(max(At)/2));
fwhmi=length(fwhmi);
dw=1/l/dt*2*pi;
w=(-1*l/2:1:l/2-1)*dw;
At=fftshift(At);
 w=fftshift(w);
spec=fft(fftshift(At)); %generating a pulse spectrum
for j=h:h:X
spec=spec.*exp(-A*(h/2)+i*B2/2*w.^2*(h/2)) ; % calculation of linear part of NLSE
f=ifft(spec);
f=f.*exp(i*gamma*((abs(f)).^2)*(h)); %calculating nonlinear part of NLSE
spec=fft(f);
spec=spec.*exp(-A*(h/2)+i*B2/2*w.^2*(h/2)) ; % NLSE calculation
end
f=ifft(spec); % pulse propagation
output_pulse(q,:)=abs(f); %preserving output pulse
fwhmo=find(abs(f)>abs(max(f)/2));
fwhmo=length(fwhmo);% finding the Full width half maximumof output signal
pbratio(q)=fwhmo/fwhmi; %finding pulse brodening ratio
```

```matlab
q=q+1;
end

figure(2);
mesh(output_pulse(1:1:q-1,:)); %output spectrum
title('Pulse Evolution');
xlabel('Time'); ylabel('distance in km'); zlabel('amplitude');

figure(3)
plot(pbratio(1:1:q-1),'b'); %pulse broadening plot
xlabel('distance in km');
ylabel('Pulse broadening ratio');
grid on;
hold on;

figure(4);
waterfall(output_pulse(1:1:q-1,:)); %output spectrum
title('Pulse Evolution');
xlabel('Time'); ylabel('distance in km'); zlabel('amplitude');
```